\documentclass[sigconf,natbib=true]{acmart}

\renewcommand\footnotetextcopyrightpermission[1]{} 

\AtBeginDocument{%
  }

\setcopyright{acmlicensed}
\copyrightyear{2025} 
\acmYear{2025} 
\setcopyright{cc}
\setcctype{by}
\acmConference[SIGIR '25]{Proceedings of the 48th International ACM SIGIR Conference on Research and Development in Information Retrieval}{July 13--18, 2025}{Padua, Italy}
\acmBooktitle{Proceedings of the 48th International ACM SIGIR Conference on Research and Development in Information Retrieval (SIGIR '25), July 13--18, 2025, Padua, Italy}
\acmDOI{10.1145/3726302.3730212}
\acmISBN{979-8-4007-1592-1/2025/07}

\usepackage[T1]{fontenc}
\usepackage{caption} 
\usepackage{algorithm}
\usepackage{algorithmic}
\usepackage{booktabs} 
\usepackage{multirow} 
\usepackage{subcaption}
\usepackage{graphicx} 
\usepackage{amsmath}
\usepackage{enumitem}
\usepackage{colortbl}
\usepackage{xcolor}
\usepackage{balance}

\setitemize{noitemsep}
\setlist{topsep=0pt, leftmargin=*}

\newcommand{\uls}{\begin{itemize}[leftmargin=*]}
\newcommand{\ule}{\end{itemize}}
\newcommand{\ols}{\begin{enumerate}[leftmargin=*]}
\newcommand{\ole}{\end{enumerate}}
\newcommand{\li}{\item}

\newcommand{\nv}{\cellcolor{lightgray}}

\newcommand{\para}[1]{\paragraph{\textnormal{\textbf{#1}.}}} 

\usepackage[utf8]{inputenc}

\usepackage{tcolorbox} 
\tcbuselibrary{skins}  
\usepackage{xcolor} 

\usepackage{adjustbox}
\usepackage{listings}

\usepackage{calrsfs}
\DeclareMathAlphabet{\pazocal}{OMS}{zplm}{m}{n}
\DeclareMathAlphabet{\pazobfcal}{OMS}{cmsy}{b}{n}

\newcommand{\sol}{S^{(P)}}

\newcommand{\llm}{\phi_{\text{LLM}}}

\usepackage{hyperref}
\usepackage{float}
\begin{document}


\title[Estimating the Quality of LLM-Generated Program Solutions]{In-Context Learning as an Effective Estimator of Functional Correctness of LLM-Generated Code}


\author{Susmita Das}
\affiliation{%
  \institution{University of Glasgow}
  \city{Glasgow}
  \country{United Kingdom}}
  \orcid{0009-0002-6518-6099}
\email{s.das.2@research.gla.ac.uk}

\author{Madhusudan Ghosh}
\affiliation{%
  \institution{Indian Association for the \\ Cultivation of Science}
  \city{Kolkata}
  \country{India}}
  \orcid{0000-0002-8330-2703}
\email{madhusuda.iacs@gmail.com}

\author{Priyanka Swami}
\affiliation{%
  \institution{University of Glasgow}
  \city{Glasgow}
  \country{United Kingdom}}
  \orcid{0009-0006-9230-1397}
  \email{swamipriyanka98@gmail.com}

\author{Debasis Ganguly}
\affiliation{%
  \institution{University of Glasgow}
  \city{Glasgow}
  \country{United Kingdom}}
  \orcid{1234-5678-9012}
\email{Debasis.Ganguly@glasgow.ac.uk}

\author{Gul Calikli}
\affiliation{%
  \institution{University of Glasgow}
  \city{Glasgow}
  \country{United Kingdom}}
  \orcid{0000-0003-4578-1747}\email{HandanGul.Calikli@glasgow.ac.uk}





\begin{abstract}

When applying LLM-based code generation to software development projects that follow a feature-driven or rapid application development approach, it becomes necessary to estimate the functional correctness of the generated code in the absence of test cases. Just as a user selects a relevant document from a ranked list of retrieved ones, a software generation workflow requires a developer to choose (and potentially refine) a generated solution from a ranked list of alternative solutions, ordered by their posterior likelihoods.
This implies that estimating the quality of a ranked
list -- akin to estimating ``relevance'' for query performance prediction (QPP) in IR -- is also crucial for generative software development, where quality is defined in terms of ``functional correctness''. In this paper, we propose an in-context learning (ICL) based approach
for code quality estimation. Our findings demonstrate that providing few-shot examples of functionally correct code from a training set enhances the performance of existing QPP approaches as well as a zero-shot-based approach for code quality estimation.

\end{abstract}



\maketitle              


\footnotetext{
Copyright is held by the owner/author(s). Publication rights licensed to ACM.  
This is the author's version of the work. It is posted here for your personal use. Not for redistribution.  
The definitive version was published in Proceedings of SIGIR '25, \url{https://doi.org/10.1145/3726302.3730212}.
}

\newboolean{showcomments}
\setboolean{showcomments}{true} 
\ifthenelse{\boolean{showcomments}}
{
	\newcommand{\nb}[3]{
		{\colorbox{#2}{\bfseries\sffamily\scriptsize\textcolor{white}{#1}}}
		{\textcolor{#2}{$\blacktriangleright$\textsf\small{#3}$\blacktriangleleft$}}}
	 \newcommand{\version}{\emph{\scriptsize$-$Id$-$}}
}{
	\newcommand{\nb}[3]{}
} 


\newcommand{\DG}[1]{\nb{Debasis:}{blue}{#1}}
\newcommand{\GC}[1]{\nb{Gul:}{orange}{#1}}
\newcommand{\SD}[1]{\nb{Susmita:}{red}{#1}}
\newcommand{\C}[1]{\nb{Comments:}{magenta}{#1}}






\section{Introduction}
Large language models (LLMs) can model semantics in an abstract and general manner without any task-specific training \cite{radford,gpt3,arora2022ask,weidinger2022taxonomy},
making them effective zero-shot predictors for a wide range of different downstream tasks, such as assessing reviews \cite{mysore2023large}, answering questions \cite{li-etal-2023-shot}, recommending relevant documents \cite{pradeep2023does} etc.

\begin{figure*}[t]
\centering
\includegraphics[width=0.72\textwidth]{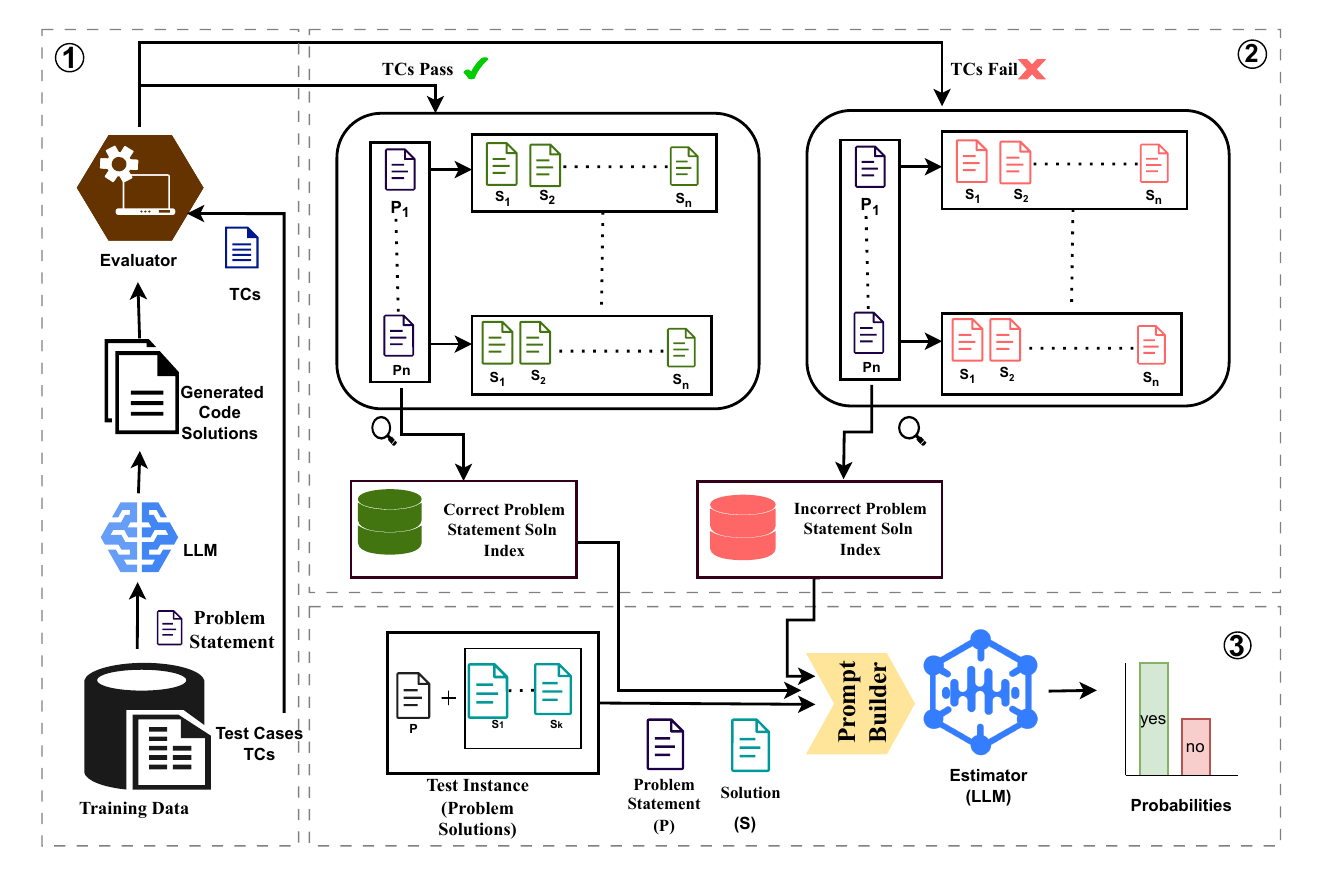}

\caption{Schematic workflow of our proposed few-shot code quality estimation: \textcircled{1} and \textcircled{2} show how training data is prepared offline by first generating solution lists for each problem and then executing test cases (TCs) on these solutions to construct two separate indexes for correct and incorrect solutions.
\textcircled{3} shows how problem-solution pairs that are most similar to the current input are retrieved and then included as ICL examples to facilitate prediction of code quality.}


\label{fig:proposed-method}
\end{figure*}


In the context of software development, LLMs have been shown to be a valuable resource for a wide range of software development tasks, such as generating a code snippet from a natural language problem description \cite{athiwaratkun2023multilingualevaluationcodegeneration,yin2022naturallanguagecodegeneration}, or complete a partially written software code to accomplish a specific task \cite{chen2021evaluatinglargelanguagemodels,wu2024repoformerselectiveretrievalrepositorylevel,guo2023longcoderlongrangepretrainedlanguage}, or even predict the quality of the generated code \cite{zhuo2024icescore}.
Such automatically generated code by LLMs can prove to be particularly beneficial under a ``human-in-the-loop'' setting, e.g., 
inexperienced software developers may find it easier to post-edit the code generated for a particular functionality without writing code from scratch \cite{wu2024repoformerselectiveretrievalrepositorylevel}, or this may help to reduce the code review time among software development teams \cite{wu2024autoapieval,generative_ai_software_dev,llm_code_meets_dev_process}.
On the flip side, LLM-generated software code may lead to risks, such as errors in terms of functional correctness \cite{dinh2023largelanguagemodelscode},
security vulnerabilities \cite{asare2024githubscopilotbadhumans}, and integration issues due to their limited understanding of a programming context \cite{rognerud2009challenges}. While unit tests can be applied in test-driven development (TDD) \cite{karac2019controlled} to check the functional correctness of code, one of the restrictions is that the test-cases need to be written before the actual implementation of a problem description.

In such situations, estimating the functional correctness of code without the presence of any unit test-cases can potentially ensure an effective and efficient software development life-cycle \cite{schäfer2023empiricalevaluationusinglarge}.

\begin{figure}[t]
\centering
\begin{adjustbox}{width=.75\columnwidth}
\begin{tcolorbox}
\small
You are an experienced software engineer. Your task is to check the functional correctness of code for the given problem statement. Generate `yes' if the code is functionally correct (i.e., code meets the problem's requirements), otherwise generate `no'.
\end{tcolorbox}
\end{adjustbox}
\includegraphics[width=.75\columnwidth]{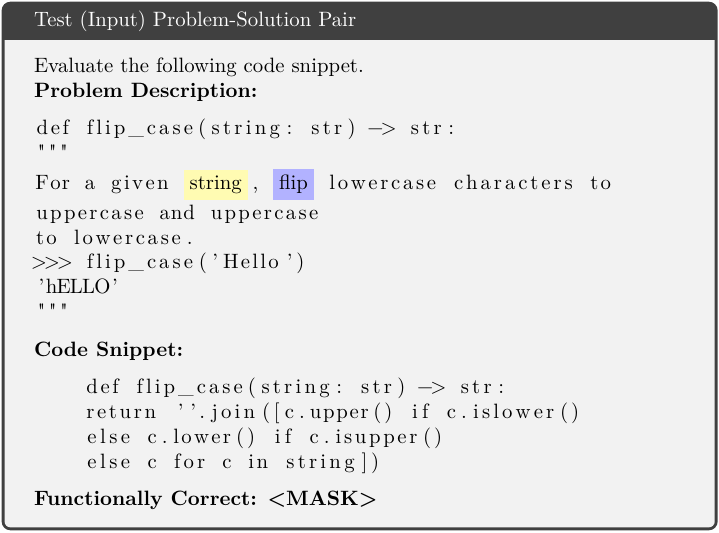}
\caption{Prompt instruction for code quality estimation with a sample input problem-solution pair.}
\label{fig:preamble}
\end{figure}

A common workflow in LLM-based software code generation involves a user (software developer) prompting an LLM to generate a number of alternative solutions ranked by the posterior likelihoods of generation. As a subsequent step, a solution from the list is eventually selected by a manual choice of what \textit{appears to be correct}. Drawing an analogy between ``\textit{functional correctness}'' and ``\textit{relevance}'', this is similar to the a user's task of finding the first relevant document by scanning a ranked list of documents. Obviously, as per the analogy, the tasks become more efficient if relevant documents are retrieved towards the top ranks, or correct solutions appear towards the top of the list - or in other words - an LLM's generation posteriors correlate well with correctness.

A code quality estimator should, therefore, seek to estimate the correlation between the posteriors of the solutions and the correctness of each, i.e., estimate if correct solutions appear before the incorrect ones in a generated list of solutions. Again, drawing the IR analogy, this task is somewhat similar to the standard task of query performance prediction (QPP) \cite{uef_kurland_sigir10,kurland_tois12,DBLP:conf/ecir/GangulyDMG22,DBLP:journals/tois/DattaGMG23} in IR - which involves estimating the quality of a ranked list of documents retrieved for a query - a high quality indicating that a relevant document is retrieved at a better rank than a non-relevant one.

This paper leverages the semantic capabilities of LLMs for the \textbf{automatic detection of code quality}. While prior research \cite{zhuo2024icescore} has demonstrated the effectiveness of zero-shot LLM inference as a code quality estimator, a key limitation of this approach is that it does not utilize labeled examples. We hypothesize that incorporating labeled examples can enhance predictive performance, as has been widely observed in other natural language processing (NLP) tasks, where in-context learning (i.e., the inclusion of labeled examples) consistently outperforms zero-shot inference \cite{li-etal-2022-encoder,li-etal-2023-shot,tang-etal-2023-context,mysore2023large,pornprasit2024fine}.
%



\section{Proposed Few-shot Code Quality Estimator} \label{sec:methodology}
Figure \ref{fig:proposed-method} shows a schematic overview of our proposed approach. In this section, we describe the details of each step involved.

\begin{figure}[t]
\centering
\includegraphics[width=.75\columnwidth]{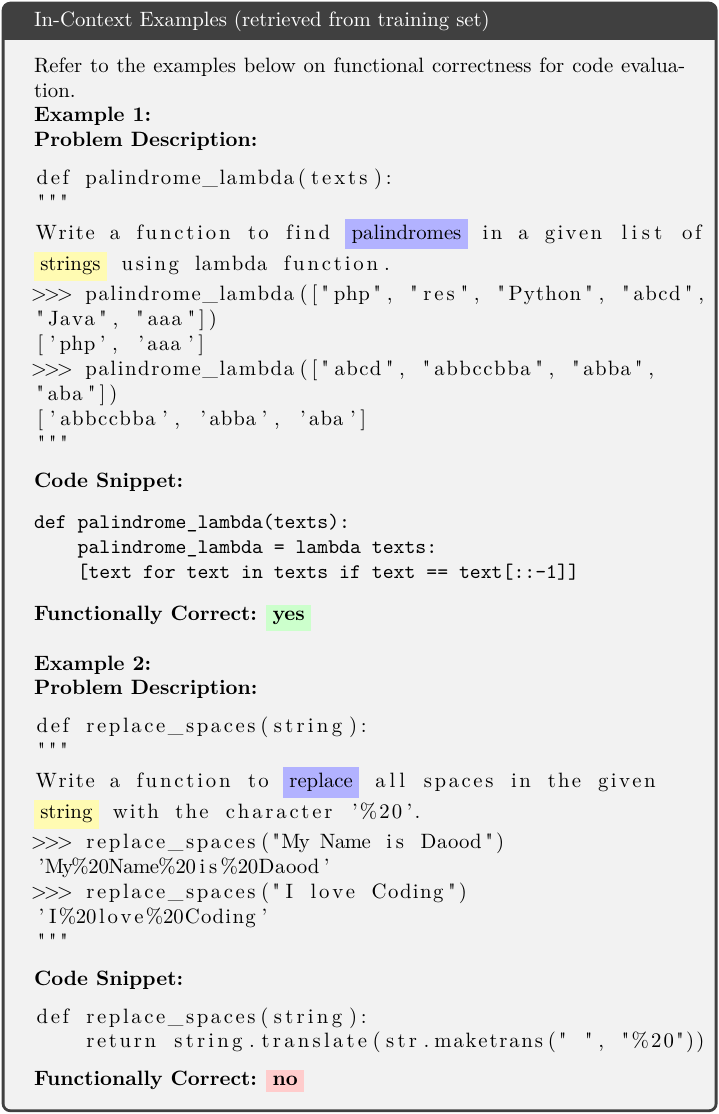}
\caption{Sample data for few-shot code quality prediction. Semantically similar words in the input and the examples are highlighted with identical colors.
\label{fig:fewshot-prompt}
}
\end{figure}

\para{Code Generation}
The input to a code quality estimator is a list of generated code solutions obtained as an output from an LLM. The input for the code generation task is a natural language description of a problem task $P$ with partially written code, such as function prototypes (see Figure \ref{fig:fewshot-prompt} for an example problem description). Formally, $\llm: I, P \mapsto \sol = \{\sol_1,\ldots,\sol_n\}$,   
where $I$ denotes a prompt instruction , $P$ the problem statement (i.e., the task to be solved), and $\llm$ denotes the frozen parameters of an LLM.

\para{Few-shot Prediction}
To estimate the quality of these generated solutions $S^{(P)}$, we propose to employ another LLM (may also be the same LLM that was used for code generation). In contrast to a 0-shot estimation as reported in \cite{zhuo2024icescore}, we propose a few-shot extension that leverages additional contextual information to control this prediction. This additional context comprises of examples of the form of problem and generated solution pairs retrieved from a training set, where the `label' for each pair indicates if the generated solution is a functionally correct implementation of the problem of the example pair, e.g., see \textcircled{2} in Figure \ref{fig:proposed-method}. The label is computed based on the availability of the test cases, e.g., see \textcircled{1} in Figure \ref{fig:proposed-method}.

Formally speaking, the code quality estimation model is a function of the form:
$\theta_{\text{LLM}}: P, \sol_i,\pazocal{C}_k(\sol_i) \mapsto \mathbb{R}$,
where $\sol_i \in \sol$ is the $i^{\text{th}}, (i=1,\ldots,n)$ solution from the list of generated solutions $\sol$, and $\pazocal{C}_k(\sol_i)$ denotes a $k$-neighborhood of similar problem-solution example pairs from a training set with available test cases, as shown in \textcircled{2} of Figure \ref{fig:proposed-method}.

\begin{table}[t]
\centering
\caption{A summary of dataset with code generation accuracy reported for two LLMs - CodeStral-22B (CS) and CodeLlama-7B (CL).
The four different evaluation settings are:
a) \textit{I} (In-domain) - same LLM for code quality prediction and code generation of training examples, b) \textit{O} (Out-domain) - different LLMs for code generation and code quality prediction, c) \textit{ML} (mono-lingual) - quality prediction for code in the same language as that of the training examples, d) \textit{XL} (cross-lingual) - quality prediction on code which in a language different from the training examples.
}
\label{tab:data}
\small 
\begin{tabular}{@{}llccccc@{}}
\toprule
\multicolumn{3}{c}{} & \multicolumn{2}{c}{Accuracy} & \multicolumn{2}{c}{Evaluation Types}\\
\cmidrule{4-5}
\cmidrule{6-7}
Mode & Dataset & $|P|$ &CS & CL & CS & CL \\
\midrule
Train & MBPP (90\%) & 877 & \multirow{2}{*}{49.61} &  \nv & \nv & \nv\\
Dev & MBPP (10\%) & 97 & &  \nv & \nv & \nv\\
\cmidrule{2-5}
\multirow{2}{*}{Test} & HEval & 164 & 51.35 & 22.51 & ID, ML & OD, ML  \\
& MBJP & 974 & 40.64 & 22.90 & ID, XL & OD, XL \\
\bottomrule
\end{tabular}
\end{table}

\para{Few-shot Similarity Function}

Retrieval of the few-shot problem-solution examples depends on how a similarity function $\sigma$ is defined between the problem-solution pairs, i.e., $\sigma: X \times X \mapsto \mathbb{R}$, where $X=(P, S)$ represents a problem-solution pair. We define $\sigma$ as a linear combination of field-based similarity measures \cite{DBLP:conf/cikm/RobertsonZT04}  - the problem and the solution being the two respective fields, i.e.,
\begin{equation}
\sigma( (P_{\text{test}}, S_{\text{test}}), (P, S)) = \alpha (e_{P_{\text{test}}}\cdot e_P) + (1-\alpha) (e_{S_{\text{test}}}\cdot e_S),
\label{eqn:similarity_function}
\end{equation}
where $(P, S)$ is a problem-solution tuple from the training set, $P_{\text{test}}, S_{\text{test}}$ is the input problem-solution pair for quality estimation, $e_Z$ denotes a dense embedded vector of the text $Z$, and $\alpha \in [0,1]$ is a parameter that controls the relative importance of similarities between the input problem with the training set ones, vs. those between the input solution and the training set ones.
As shown in Figure \ref{fig:proposed-method} we construct a balanced context comprising $k$ examples for both correct and incorrect solutions.

Specifically, we use CodeBERT \cite{feng2020codebertpretrainedmodelprogramming} as the dense embedding function for computing the similarities in Equation \ref{eqn:similarity_function}. We also experiment with three different settings for $\alpha \in \{1, 0.5, 0\}$, which means that we assign equal importance to both the problem-problem and the solution-solution matches for constructing the few-shot examples, in addition to the degenerate cases of considering only the problem or only the solution for finding the few-shot context. We name these three variants of our proposed approach as \textbf{FS-P}, \textbf{FS-S}, and \textbf{FS-PS}, respectively denoting few-shot with similarity functions based on either problems or solutions alone or as a combination.

\para{LLM-Prompting Details}
As the predictor LLM $\theta_{\text{LLM}}$, we employ a relatively small-sized LLM comprising of 7B parameters; specifically, we use the instruction-tuned version of CodeLlama-7B.
The instruction preamble is shown in Figure \ref{fig:preamble}, whereas sample inputs along with example problem-solution pairs are shown in Figure \ref{fig:fewshot-prompt}.
The final estimation likelihood is obtained by the posterior probability of generating a `yes' token relative to a `no' token, i.e., $P(\text{`yes'}|\theta_{\text{LLM}})/(P(\text{`yes'}|\theta_{\text{LLM}}) + P(\text{`no'}|\theta_{\text{LLM}}))$.

\section{Experiment Setup}  

\begin{table}[t]
\centering
\caption{Code quality estimation results for different settings.
For each few-shot (proposed method), we show the dev-set tuned values of $k$ (\#examples).
\label{tab:code_quality_estimation}
}


\small
\begin{tabular}{c@{~~~~}|c@{~~~~}| c@{~~}c@{~~}c@{~~}c@{~~}|c@{~~}c@{~~}c@{~~}c}
\toprule
\multicolumn{6}{r}{Global nDCG} & \multicolumn{4}{c}{Local nDCG} \\
\cmidrule(r){3-6} \cmidrule(r){7-10}    
\multicolumn{4}{r}{CS (ID)} & \multicolumn{2}{c}{CL (OD)} & \multicolumn{2}{c}{CS (ID)} & \multicolumn{2}{c}{CL (OD)}  \\
\cmidrule(r){3-4} 
\cmidrule(r){5-6}
\cmidrule(r){7-8}
\cmidrule(r){9-10}
\multicolumn{2}{c}{} & HEval & MBJP & HEval & MBJP & HEval & MBJP & HEval & MBJP \\
Model & $k$ & (ML) & (XL) & (ML) & (XL) & (ML) & (XL) & (ML) & (XL) \\
\midrule




ELS & n/a  & .903 &  .885 & .786 & .823 & .640 &  .518 & .260 & .254 \\
TLS & n/a & .902 &  .887 & .708 & .785 & .638 &  .523 & .265 & .256 \\
ZS & n/a & .904 &  .890 & .797 & .880 & .635 &  \textbf{.570} & .317& .269 \\
\midrule

FS-PS & 4 & \textbf{.915} & \textbf{.906} & \textbf{.880} & .873 & \textbf{.658} & .549 & \textbf{.318} & .271 \\

FS-P & 4 & .891 & .904 & .842 & .870 & .649  & .545 & .317 & \textbf{.275} \\

FS-S & 3 & .910 & .903 & .875 & \textbf{.887} & .658 & .542 & .312 & .270 \\


\bottomrule
\end{tabular}

\end{table}

\para{Research Questions}
We conduct experiments to validate the following research questions.
\uls
\li \textbf{RQ-1}: Do examples improve code quality prediction over 0-shot?

\li \textbf{RQ-2}: Can an LLM with smaller number of parameters be used to predict the quality of code generated by an LLM of a different family and large parameter size?

\li \textbf{RQ-3}: Should the ICL examples be obtained by considering similarities between problems or solutions alone or as a combination?

\li \textbf{RQ-4}: Does the few-shot method generalize across a different programming language?
\ule

\begin{figure*}[t]
\centering
\begin{subfigure}[t]{0.40\columnwidth}
    \centering
    \includegraphics[width=\linewidth]{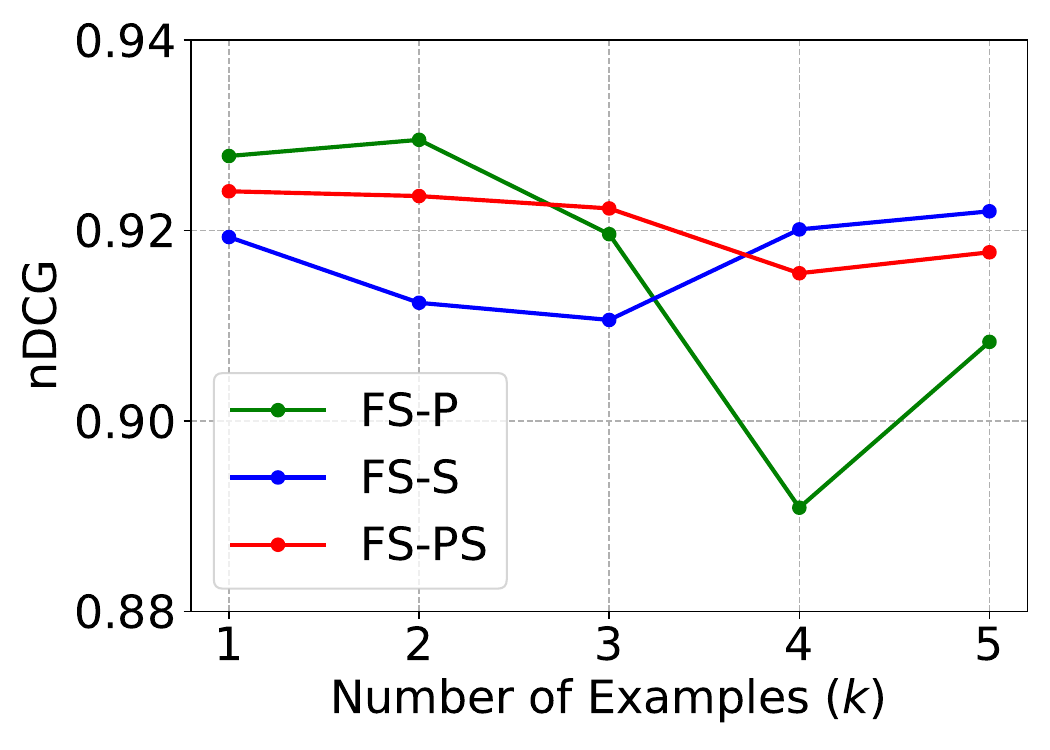}  
    \caption{G-nDCG on HEval (CS)}
    \label{fig:codestral}
\end{subfigure}
\begin{subfigure}[t]{0.40\columnwidth}
    \centering
    \includegraphics[width=\linewidth]{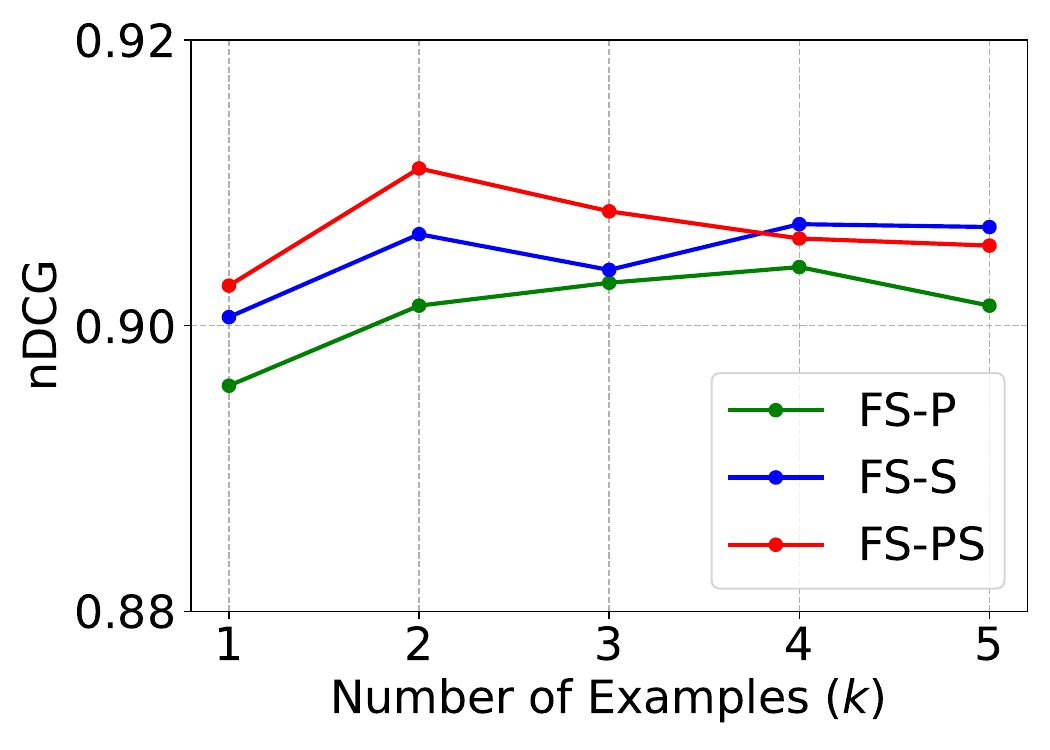}
    \caption{G-nDCG on MBJP (CS)}
    \label{fig:codellama}
\end{subfigure}
\begin{subfigure}[t]{0.40\columnwidth}
    \centering
    \includegraphics[width=\linewidth]{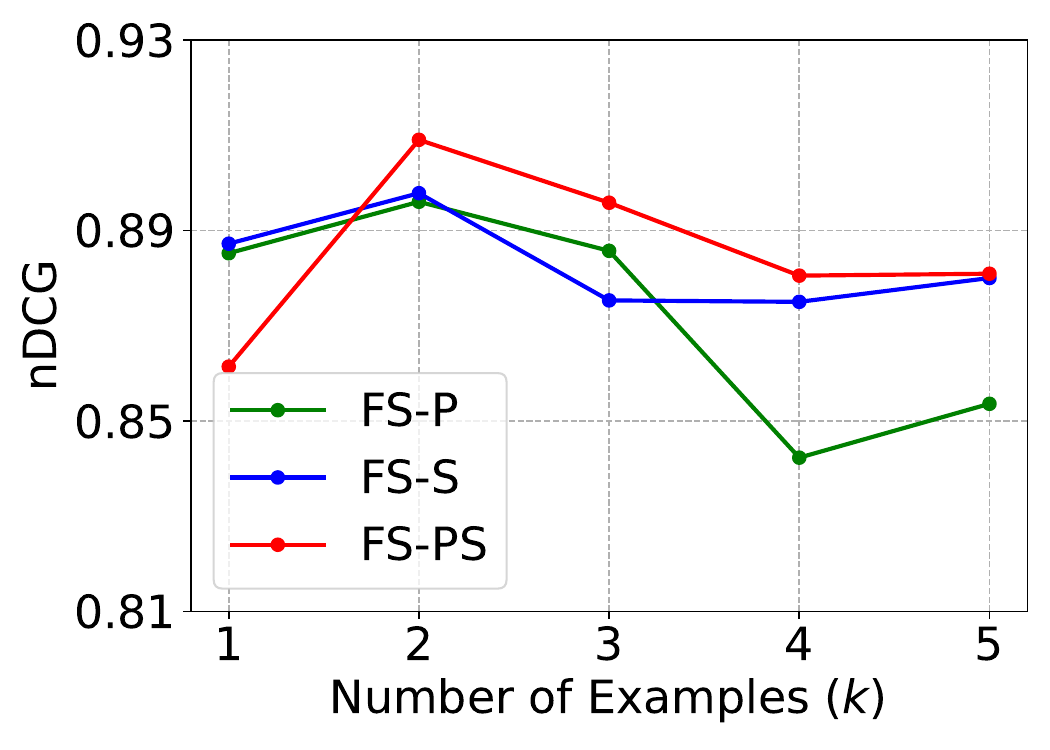}
    \caption{G-nDCG on HEval (CL)}
    \label{fig:codestral-mbjp}
\end{subfigure}
\begin{subfigure}[t]{0.40\columnwidth}
    \centering
    \includegraphics[width=\linewidth]{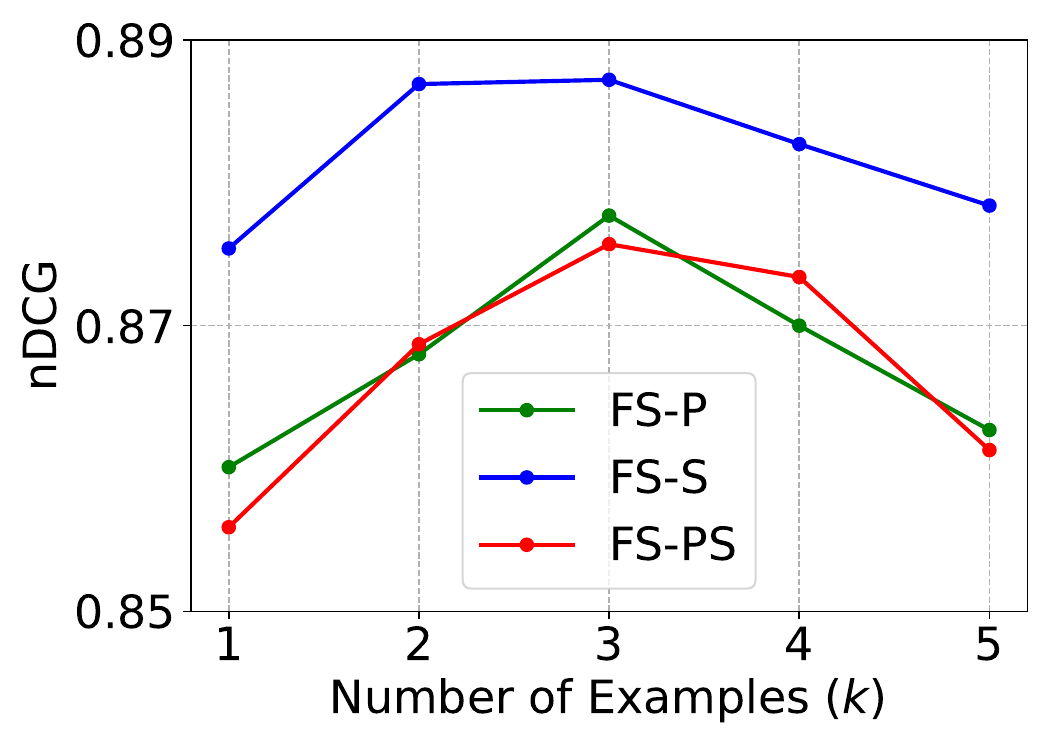}
    \caption{G-nDCG on MBJP (CL)}
    \label{fig:codellama-mbjp}
\end{subfigure}

\begin{subfigure}[t]{0.40\columnwidth}
    \centering
    \includegraphics[width=\linewidth]{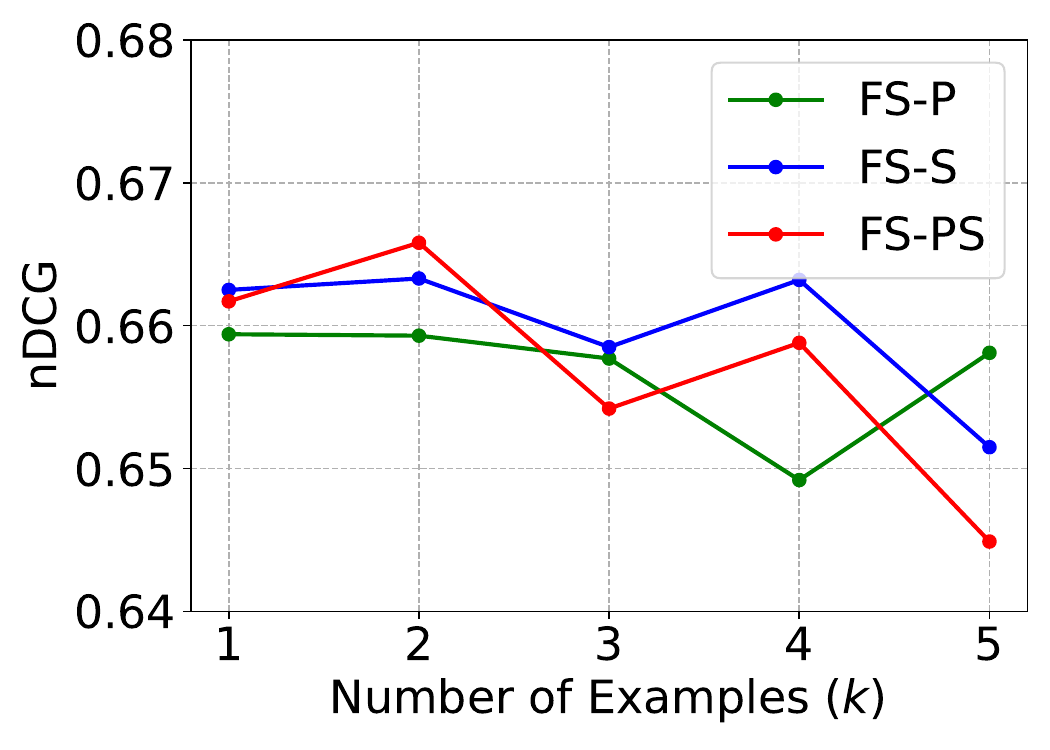}  
    \caption{L-nDCG on HEval (CS)}
    \label{fig:local-codestral}
\end{subfigure}
\begin{subfigure}[t]{0.40\columnwidth}
    \centering
    \includegraphics[width=\linewidth]{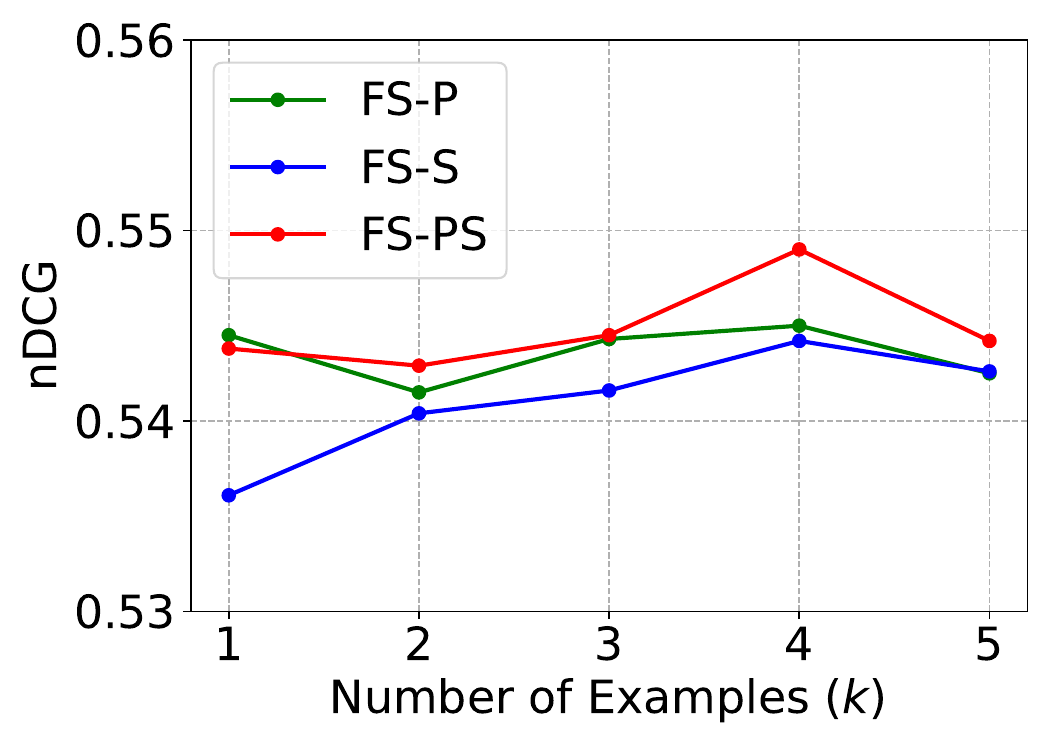}
    \caption{L-nDCG on MBJP (CS)}
    \label{fig:local-codestral-mbjp}
\end{subfigure}
\begin{subfigure}[t]{0.40\columnwidth}
    \centering
    \includegraphics[width=\linewidth]{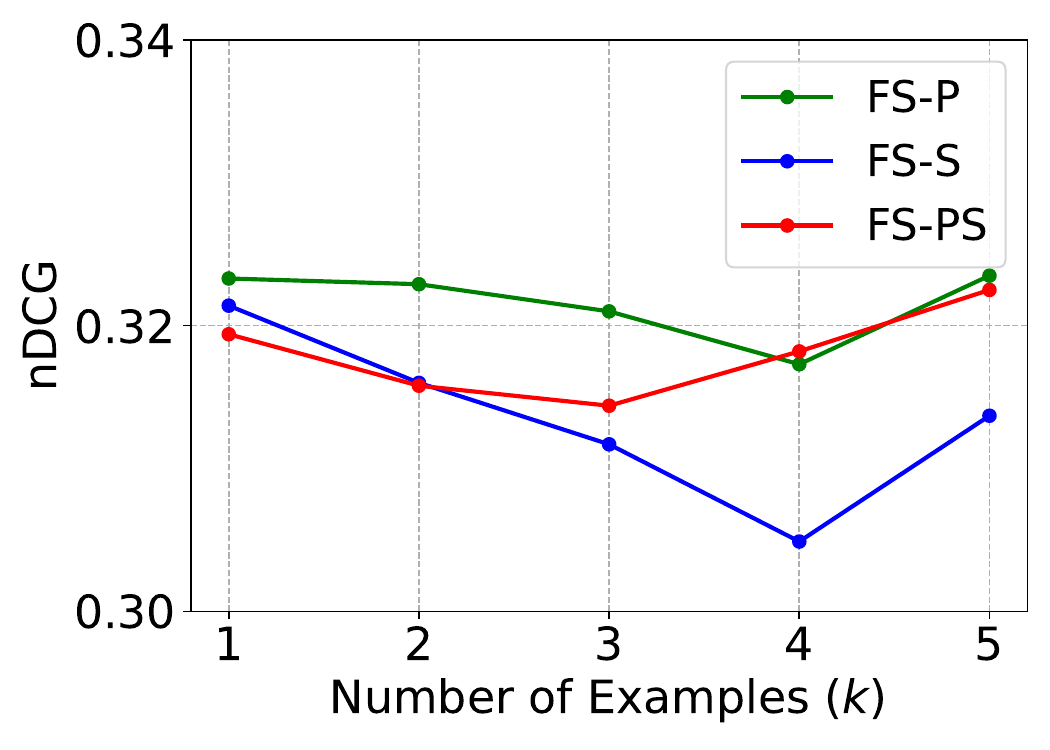}
    \caption{L-nDCG on HEval (CL)}
    \label{fig:local-codelama-heval}
\end{subfigure}
\begin{subfigure}[t]{0.40\columnwidth}
    \centering
    \includegraphics[width=\linewidth]{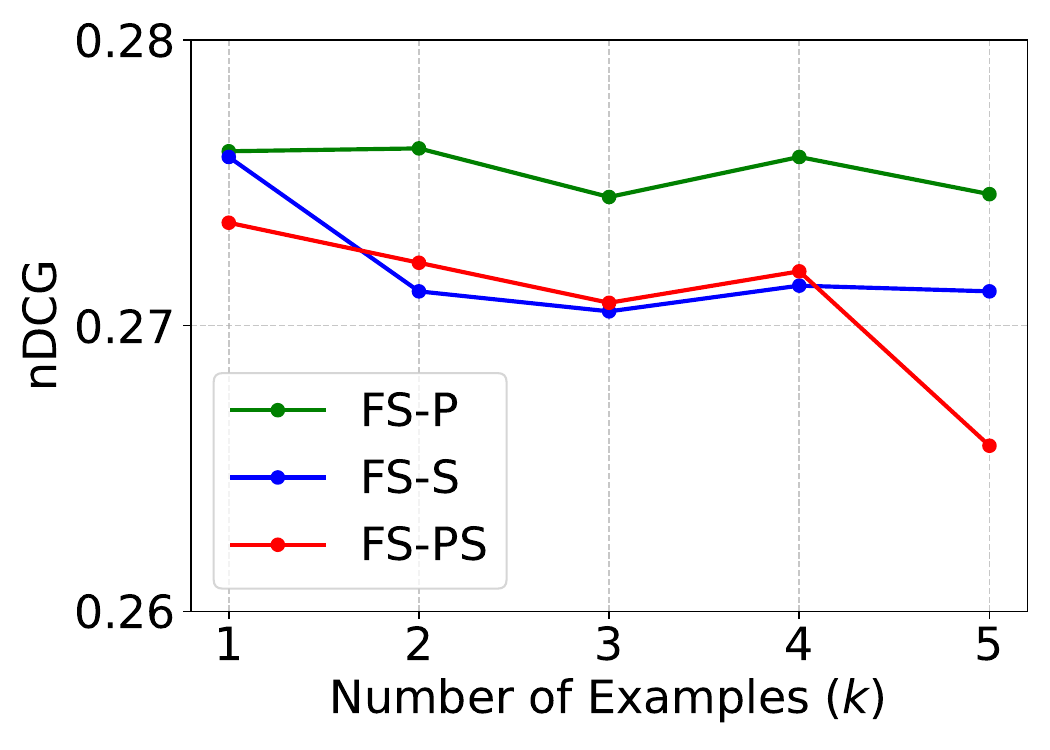}
    \caption{L-nDCG on MBJP (CL)}
    \label{fig:local-codellama-mbjp}
\end{subfigure}
\caption{Effect of variations in $k$ on the global and local tasks - respectively, ranking all problem-solution pairs, vs. ranking a list of 10 solutions for each problem for the two different test sets.
\label{fig:ndcg_comparison}
}
\end{figure*}

\para{Datasets}
For our experiments we use the standard benchmark code generation Python datasets - HumanEval \cite{chen2021evaluatinglargelanguagemodels} and MBPP \cite{austin2021programsynthesislargelanguage}.
We use MBPP as our training set for ICL examples (see Figure \ref{fig:proposed-method}). 
For in-domain testing, we use HumanEval (HEval), whereas for out-domain (cross-lingual) evaluation, we use the MBJP dataset, which is the Java version of MBPP.
Table \ref{tab:data} provides a summary of the dataset statistics.

As code generation models, we employ Codestral-22B and CodeLlama-7B. As examples of functional correctness of generated solutions for each problem in the MBPP training dataset, we select the Codestral-22B generated solutions as per the hypothesis that the solutions generated by a larger model may generalize well for MBJP. In particular, we generated 10 solutions for each problem in MBPP.
%
Table \ref{tab:data} shows the four different evaluation types corresponding to the differences in the generation and the prediction models (ID or OD), and to the differences in the language of the examples and the input (ML or XL).

\para{Baselines}
As baselines, we employ the following approaches.

\uls
\li \textbf{TLS} \cite{zhou-etal-2023-codebertscore}: Computes the inter similarity between CodeBERT embeddings \cite{feng2020codebertpretrainedmodelprogramming} of two solutions $(S,S')$ of a problem $P$ with the assumption that higher similarities is likely to indicate a more effective  solution similar to CodeBERTScore \cite{zhou-etal-2023-codebertscore}.

\li \textbf{ELS}: A simplified version of TLS, which predicts the similarity between the [CLS] representations between two solutions $(S,S')$ of a problem $P$.

\li \textbf{ZS} \cite{zhuo2024icescore}: A 0-shot version of our proposed few-shot based predictor, which simply uses the prompt instruction (Section \ref{sec:methodology}) and the current input $(P,S)$, i.e., problem-solution pair, (see Figure \ref{fig:fewshot-prompt}) to yield a posterior likelihood for functional correctness of $S$.

\ule

\para{Evaluation Metrics and Parameter Settings}

A predictor can be considered effective if it predicts a high score for a correct solution and a low score for an incorrect one  such that functionally correct solutions appear toward the top of the ranking. 
We evaluate the quality of this predictor induced ranking of problem-solution pair by means of the standard ranking measure of nDCG \cite{ndcg}. The ground-truth label is a Boolean indicator of whether solution passes all the test cases for the problem.

Following prior works \cite{zhou-etal-2023-codebertscore,zhuo2024icescore}, we evaluate the effectiveness of code quality estimation models at two levels -- local (L-nDCG) and global (G-nDCG). In L-nDCG, the 10 solutions for a particular problem is to be ranked in a decreasing order of likelihood of functional correctness, whereas in G-nDCG, a set of problem-solution pairs comprised of several problem instances need to be ranked globally. G-nDCG, thus evaluates how effectively an estimator is able to compare across solutions for different problem instances. G-nDCG is a single measure and L-nDCG is averaged over all problems.


To tune the hyper-parameter $k$ (number of correct and incorrect code solutions for examples) we use the 10\% samples from MBPP as the development set (as shown in Table \ref{tab:code_quality_estimation}). We optimize $k$ on the dev set for G-nDCG measure, and use this setting for all the four different test scenarios (see Table \ref{tab:data}). 
Our implementation is publicly available for research  purposes. \footnote{\href{https://github.com/Susmitacse/ICL-Code-Quality-Estimation.git}{https://github.com/Susmitacse/ICL-Code-Quality-Estimation.git}}

\section{Results} \label{sec:results}

\para{Main findings (Table \ref{tab:code_quality_estimation})}
\textit{First}, in relation to RQ-1, we observe that ICL outperforms 0-shot baselines, which can be seen by comparing the results of the lower half of the table (our proposed few-shot) with the upper half (the baselines). The improvements are consistent for both local and global settings of the predictor. 
\textit{Second}, for RQ-2, we observe that the Codestral (CS) examples does help to improve the effectiveness of code quality prediction (compare the OD results of the baselines vs. the FS methods).
\textit{Third}, for RQ-3, we observe that among the different options for example selection, considering both problems and solutions for computing the neighborhood of similar examples (i.e., the FS-PS variant) mostly outperforms the problem-only (FS-P) and the solution-only variants (FS-S). For a cross-lingual setup, it turns out that FS-PS is not the best approach. This is most likely because cross-language similarities between solutions are less reliable.
\textit{Finally}, for RQ-4, similar to the OD setup, we observe that the prediction effectiveness of Java code improves with Python examples of functional correctness, as seen from the evaluation measures corresponding to the XL settings in Table \ref{tab:code_quality_estimation}.

\para{Sensitivity to \#examples}
We also conduct a comparative post-hoc analysis of the effect of the number of examples on the different few-shot approaches. The plots in Figure \ref{fig:ndcg_comparison} reveal that FS-PS is mostly better than the other variants. However, as expected, in a (OD, XL) setup (different LLM, different language) is worse than FS-P and FS-S. While the former (Figure \ref{fig:ndcg_comparison} h) is better for the local task, the latter is better for the global one (Figure \ref{fig:ndcg_comparison} d).

\para{Concluding Remarks}
We demonstrated that in-context learning with relatively small-sized LLMs can lead to effective code quality estimation thus indicating its potential application in feature-driven software development.
As a part of future work, we plan explore the effectiveness of code quality estimation models for a more challenging setup, i.e., when only a small proportion of test cases, or none, is available to be used during training or evaluation. Furthermore, we also intend to investigate the code quality in terms of security, maintainability, or
efficiency.



\bibliographystyle{ACM-Reference-Format}
\balance

\bibliography{refs}





\end{document}